\documentclass[pra,aps,amssymb,nofootinbib,twocolumn]{revtex4}
\usepackage{amsmath}
\usepackage{amssymb}
\usepackage{amsthm}
\usepackage{amsfonts}
\usepackage{algorithmic}
\usepackage{enumerate}
\usepackage{latexsym}
\usepackage[dvips]{graphicx}

\newcommand{\beq}{\begin{equation}}
\newcommand{\eneq}{\end{equation}}
\newcommand{\beqs}{\begin{equation*}}
\newcommand{\eneqs}{\end{equation*}}

\input{epsf}

\begin{document}

\tolerance 10000

\title{On Superfluidity and Suppressed Light Scattering in BECs}

\author { Zaira Nazario$^\dagger$ and
David I. Santiago$^{\dagger, \star}$ }

\affiliation{ $\dagger$ Department of Physics, Stanford
University,Stanford, California 94305 \\ 
$\star$ Gravity Probe B Relativity Mission, Stanford, California 94305}

\begin{abstract}
\begin{center}

\parbox{14cm}{ We show that the suppression of light scattering off a
Bose Einstein Condensate is equivalent to the Landau argument for
superfluidity and thus is a consequence of the {\it Principle of
Superfluidity}. The superfluid ground state of a BEC contains
nonseparable, nontrivial correlations between the bosons that make up
the system, i. e., it is entangled. The correlations in the ground
state entangle the bosons into a coherent state for the lowest energy
state. The entanglement is so extreme that the bosons that make up the
system cannot be excited at long wavenumbers. Their existence at low
energies is impossible. Only quantum sound can be excited, i.e. the
excitations are Bogolyubov quasiparticles which do not resemble free
bosons whatsoever at low energies. This means that the system is
superfluid by the Landau argument and the superfluidity is ultimately
the reason for suppressed scattering at low wavelengths. }

\end{center}
\end{abstract}

\date{\today}

\maketitle

The advent of Bose Einstein Condensates (BECs) in
1995\cite{bec1,bec2,bec3} was followed by an explosion of experimental
activity. Among the experimental efforts figure studies of spontaneous
emission and light scattering of these superfluid systems by
W. Ketterle and collaborators\cite{kett1}. They found that light
scattering off BECs is suppressed to negligible values at long
wavelengths. This seems surprising at first since one can naively
expect the typical $n + 1$ bosonic enhancement\cite{kett2} as observed
in their emission experiments\cite{kett1}. They have attributed the
observed suppression of light scattering in BECs to destructive
quantum interference effects\cite{kett2}. While their argument is
certainly true, it is the purpose of this brief note to point out that
the more universal reason that makes their argument, and hence their
results, true is the {\it Principle of Superfluidity}.

BECs are systems of bosonic atoms with short range repulsive
interactions among them, as all atoms will repel each other when
placed close enough together. At low enough temperatures, more
importantly than being Bose condensed, the repulsion correlates
particles with momentum ${\bf k}$ and $-{\bf k}$ into a coherent ground
state. Such a repulsion endows the ground state with the necessary
rigidity that makes superflow possible. Such a rigidity impedes the
existence of soft quasiparticles excitations, allowing only quantum
sound as a low energy excitation. Hence the event of light scattering
and its concomitant excitations, which will necessarily disrupt the
superfluid, will be suppressed by virtue of the Principle of
Superfluidity.

An apt Hamiltonian that encapsulates the universal physics of Bose
Einstein Condensates is the Bose Hubbard model.

\begin{multline}
\mathcal{H} = \sum_{\bf k} \epsilon_{\bf k} a_{\bf k}^\dagger a_{\bf k} - \mu 
\sum_{\bf k} a_{\bf k}^\dagger a_{\bf k} \\
+ \frac{U_0}{2} \sum_{{\bf k}, 
{\bf k_2}, {\bf q}} a_{{\bf k} + {\bf q}}^\dagger 
a_{{\bf k}_2 - {\bf q}}^\dagger a_{\bf k} a_{{\bf k}_2}
\end{multline}

\noindent where $a_{\bf k}^\dagger$ and $a_{\bf k}$ are bosonic
creation and annihilation operators, $\epsilon_{\bf k} = \hbar^2 k^2 /
2 m$ is the kinetic energy or dispersion of the noninteracting bosons,
$\mu$ is a chemical potential introduced to conserve particles when we
make the Bogolyubov approximation, and U represents the interaction
between bosons.

We consider first ground state properties. We suppose that the
parameters of our Hamiltonian is such that the ground state of the
system is a Bose superfluid and not any other competing phase such as
a Mott insulator. A precondition on the superfluid system is for it to
be a quantum fluid. That is, at low temperatures it is Bose condensed,
with a macroscopic number of bosons occupying the zero momentum
state. This is why He$^4$ has the superfluid transition at the Bose
condensation temperature. Under macroscopic occupation we can treat the
operators $a_0$ and $a_0^\dagger$ as c numbers and replace them by
$\sqrt{N_0}$, the square root of the total number of bosons in the
condensate. We separate terms corresponding to the condensate from the
interacting Hamiltonian. Higher order terms can be discarded because the
correlations in the ground state dominate at low ${\bf k}$ and make
those terms negligible. This leads us to a reduced Bogolyubov
Hamiltonian\cite{bogo}

\begin{multline} \label{redham}
\mathcal H = \frac{ 1 }{ 2 } N_0^2 U_0 + \sum_{ {\bf k} \neq 0 } 
\tilde { \epsilon_{\bf k} } \; a_{\bf k}^\dagger 
a_{\bf k} \\
+ U_0 \frac{ N_0 }{ 2 } \sum_{ {\bf k} \neq 0 } \; 
( a_{\bf k}^\dagger a_{ - {\bf k} }^\dagger + a_{ - {\bf k} } a_{\bf k} )
\end{multline}

\noindent where 

\beq \label{etilde}
\tilde { \epsilon_{\bf k} } = \frac{ \hbar^2 k^2 }{ 2 m } + N_0 U_0
\eneq

\noindent and $\mu$ has been eliminated by a variational condition on
the ground state energy with respect to number of condensate
particles, $N_0$.

The Hamiltonian in equation (\ref{redham}) is diagonalized by a
Bogolyubov transformation

\beq \label{bogo}
b_{\bf k} = u_{\bf k} a_{\bf k} + v_{\bf k} a_{ - {\bf k} }^\dagger 
\qquad b_{\bf k}^\dagger = u_{\bf k} a_{\bf k}^\dagger + v_{\bf k} 
a_{ - {\bf k} }
\eneq

\noindent by choosing 

\beq \label{uv}
u_{\bf k} = \frac{ 1 }{ \sqrt{2} } \sqrt{ \frac{ \tilde{\epsilon_{\bf
k}} } { E_{\bf k} } + 1 } \qquad v_{\bf k} = \frac{ 1 }{ \sqrt{2} }
\sqrt{ \frac{ \tilde{\epsilon_{\bf k}} }{ E_{\bf k} } - 1 } \eneq

\beq \label{energy}
E_{\bf k} = \sqrt{ \tilde { \epsilon_{\bf k} }^2 - N_0^2 U_0^2 } = 
|k| \sqrt{\frac{ \hbar^4 k^2 }{ 4 m^2 } + \frac{ \hbar^2 N_0 U_0 }
{ m }}
\eneq

\noindent where the Bogolyubov operators satisfy bosonic commutation
relations. In the long wavelength limit ($ k \rightarrow 0 $) we
arrive at a quasiparticle excitation spectrum

\beq
E_{\bf k} \simeq \hbar |k| \sqrt{\frac{ N_0 U_0 }{ m }} = 
p \sqrt{\frac{ N_0 U_0 }{ m }} 
\eneq

\noindent where $ p = \hbar |k| $. We see that the low energy
excitations have sound dispersion $ E_{\bf k} = p c_s $ with sound
speed

\beq \label{sound}
c_s = \sqrt{\frac{ N_0 U_0 }{ m }}
\eneq

This is a finite measurable quantity\cite{sound1} that indicates the
presence of correlations among the constituents of the material since
in the absence of repulsion the correlations, and with them the sound
speed, will go to zero. Such correlations stabilize persistent
currents and give rigidity to the ground state by making it
energetically expensive to excite the fundamental bosons
independently.  Instead, the excitations are collective modes and the
system superflows\cite{landau}. The ground state has acquired
rigidity.

The absence of boson repulsion, and hence rigidity, would permit
scattering processes to occur which impedes superfluidity through
dissipation. This is equivalent to Landau's argument for
superfluidity, which we will briefly sketch. A Bose fluid at zero
temperature moving with velocity $\mathbf v$ and interacting with an
external agent can exchange energy and momentum with the agent. The
fluid has initial energy $M v^2 /2$ with $M$ the total mass of the
fluid. The interaction creates a quasiparticle with energy
$\epsilon_{\mathbf p}$ {\it in the frame of} {\it the Bose fluid}. In
the lab frame the particle has energy

\beq
\tilde{\epsilon}_{\mathbf p} = \epsilon_{\mathbf p} 
+ \mathbf p \cdot \mathbf v
\eneq
  
\noindent by Galilean invariance\cite{landau}. If the fluid can lower
its energy by exciting quasiparticles it will do so no matter how mild
the interaction is, and hence it will dissipate. The final fluid
energy with a quasiparticle excited is

\beq \label{gali}
\frac{M v^2}{2} + \epsilon_{\mathbf p} + \mathbf p \cdot \mathbf v \; 
\eneq

\noindent The condition for dissipation is therefore
$\epsilon_{\mathbf p} + \mathbf p \cdot \mathbf v < 0$. The easiest
way to satisfy the condition is when $\mathbf p$ is antiparallel to
$\mathbf v$. We then obtain dissipation for

\beq
v > \left( \frac{\epsilon_{\mathbf p}}{p} \right)_{\text{min}} \; 
\eneq

\noindent For all dispersions softer than sound the minimum on the
right hand side of the last equation is zero and there is dissipation
for {\it any velocity of the fluid.} If the dispersion is at least as
stiff as sound, for velocities smaller than some critical velocity
there is no dissipation and hence superflow. We thus see that Bose
condensation is a necessary but not sufficient condition for the
existence of superfluidity. One also needs the ground state rigidity
provided by boson repulsion.

We will now move to study light scattering off the Bose Einstein
Condensate. As we will show below and had been previously shown by
Ketterle and collaborators\cite{kett2} the suppresses ed scattering
follows from destructive quantum interference that does not permit
bosons to be excited out of the fluid. Such destructive interference
will be absent if $v_{\bf k} = 0$. We point out that the speed of
sound, $c_s$, of the system will be zero {\it if and only if}
$v_{\bf k} = 0$. Therefore the suppressed scattering is a consequence
of the system being a superfluid and hence equivalent to the Landau
argument.

In order to study scattering off the superfluid we will need to
calculate perturbation matrix elements of excited states with the
ground state. The normalized BEC ground state wavefunction is

\beq \label{gs}
\qquad \quad |\Psi_0 \rangle = \prod_{\bf k} 
\frac{ 1 }{ u_{\bf k} } e^{ -(v_{\bf k} / u_{\bf k}) a_{\bf k}^\dagger 
a_{-\bf k}^\dagger} \; |0 \rangle
\eneq

\noindent In the absence of interactions, $u_{\bf k} = 1, v_{\bf k} =
0$ and we have a noncorrelated ground state. In this case, when we
excite the system by scattering, we literally pull out one of the
bosonic atoms constituting the condensate and give it nonzero
momentum. In this noninteracting and hence nonsuperfluid case,
scattering will have a typical bosonic enhancement. In the superfluid
case, which corresponds to interacting bosons, atoms with momentum
${\bf k}$ and $-{\bf k}$ are mixed so when an excitation with momentum
${\bf k}$ is created, another with momentum $-{\bf k}$ is destroyed
forming a coherent superposition.

In a scattering event, an incoming particle couples to the density of
the system. The effective scattering Hamiltonian is

\beq
\mathcal H_I = \int d^3 r \; V ( {\bf r}, t ) \rho ({\bf r}) + \text{H.C.} 
\eneq

\noindent or after Fourier transforming

\beq
\mathcal H_I = \sum_{\bf k} V_{\bf q} e^{i \omega t}
a_{{\bf k} + {\bf q}}^\dagger a_{\bf k} + \text{H.C.}
\eneq

\noindent where the energy and momentum transfered in the scattering
event are

\beq
{\bf q} = {\bf q}_f - {\bf q}_i \; \; \text {and} \; \; \omega 
= \omega_i - \omega_f
\eneq

\noindent and where we have concentrated only on one ${\bf q}$. This is
enough at first order where the amplitudes for the different momenta
add.

In order to insure adiabaticity in the interaction we introduce an
extra term in the exponential of the interaction Hamiltonian. The
interaction will be turned on at $t = - \infty$ and left on until a
time $t$. At the end of the calculations, we will take the limit $\eta
\rightarrow 0^+$.

\beq
\mathcal H_I = \sum_{\bf k} V_{\bf q} e^{i ( \omega - i \eta ) t }
a_{{\bf k} + {\bf q}}^\dagger a_{\bf k} + \text{H.C.}
\eneq

Calculating the time evolution of the Bose system in the presence of
the scattering particle by means of time dependent perturbation theory
(with the ground state $| \Psi_i \rangle = | \Psi_o \rangle$, $E_i =
0$ as the initial state) we obtain for the amplitude, after separating
the ${\bf k} = 0, -{\bf q}$ momentum states to account for the
condensate

\begin{widetext} 
\begin{multline} \label{ampl}
c_m (t) = - \frac{ V_{\bf q} \sqrt {N_0} ( \langle \Psi_m | 
a_{\bf q}^\dagger | \Psi_0 \rangle + \langle \Psi_m | a_{-\bf q} 
| \Psi_0 \rangle ) }{E_m + \hbar \omega - i \hbar \eta } \; 
e^{i(E_m + \hbar \omega - i \hbar \eta) t/ \hbar} \; - \frac { V_{\bf q} 
\sqrt {N_0} ( \langle \Psi_m | a_{\bf q} |\Psi_0 \rangle + \langle \Psi_m 
| a_{-\bf q}^\dagger | \Psi_0 \rangle ) }{E_m - \hbar \omega - i \hbar \eta } 
\; e^{i(E_m - \hbar \omega - i \hbar \eta) t/ \hbar} \\
- \sum_{ {\bf k} \neq 0, -{\bf q} } V_{\bf q} \frac{ \langle \Psi_m | 
a_{ {\bf k} + {\bf q} }^\dagger a_{\bf k} | \Psi_0 \rangle }{ E_m  
+ \hbar \omega - i \hbar \eta } e^{ i ( E_m + \hbar \omega - i \hbar \eta ) 
t / \hbar } - \sum_{ {\bf k} \neq 0, -{\bf q} } V_{\bf q} 
\frac{ \langle \Psi_m | a_{\bf k}^\dagger a_{ {\bf k} + {\bf q} } 
| \Psi_0 \rangle }{ E_m  - \hbar \omega - i \hbar \eta } e^{ i ( E_m 
- \hbar \omega - i \hbar \eta ) t / \hbar }
\end{multline}
\end{widetext}

Since the scattering particles couple to the density, the amount of scattering 
is proportional to the density component with momentum {\bf q}, the so called 
density response function

\beq \label{density}
\langle \rho_{\bf q} \rangle = \sum_m c_m \langle \Psi_0 | \rho_{\bf q} |
\Psi_m \rangle + \sum_m c_m^* \langle \Psi_m | \rho_{\bf q} | \Psi_0 
\rangle
\eneq

\noindent Notice from (\ref{bogo}) that 

\beq
a_{\bf q}^\dagger = u_{\bf q} b_{\bf q}^\dagger - v_{\bf q} b_{ - {\bf q} } 
\qquad a_{\bf q} = u_{\bf q} b_{\bf q} - v_{\bf q} b_{ - {\bf q} }^\dagger
\eneq

\noindent Using these relations together with the expression for the
density $\rho_{\bf q} = \sum_{\bf k} a_{ {\bf k} + {\bf q} }^\dagger
a_{\bf k}$ and equation (\ref{ampl}) we find

\begin{widetext}
\begin{align} \label{dens}
\langle \rho_{\bf q} \rangle &= \frac{ - V_{\bf q} N_0 
( u_{\bf q} - v_{\bf q} )^2 }{ E_{\bf q} - \hbar \omega - i \hbar \eta } 
e^{ i ( E_{\bf q} - \hbar \omega - i \hbar \eta ) t / \hbar } 
e^{ -i E_{\bf q} t / \hbar } - \frac{ V_{\bf q} N_0 
( u_{\bf q} - v_{\bf q} )^2 }{ E_{\bf q} + \hbar \omega + i \hbar \eta } 
e^{ -i ( E_{\bf q} + \hbar \omega + i \hbar \eta ) t / \hbar } 
e^{ i E_{\bf q} t / \hbar }  \nonumber \\
&- \sum_{m} \sum_{ {\bf k} \neq 0, -{\bf q} } 
\sum_{ {\bf l} \neq 0, -{\bf q} } \frac{ u_{\bf k} u_{\bf l} 
v_{ {\bf k} + {\bf q} } v_{ {\bf l} + {\bf q} } V_{\bf q} }
{ E_{ {\bf k} + {\bf q} } + E_{\bf k} - \hbar \omega - i \hbar \eta } 
e^{ i ( E_{ {\bf k} + {\bf q} } + E_{\bf k} - \hbar \omega - i \hbar \eta ) 
t / \hbar } e^{ -i ( E_{ {\bf k} + {\bf q} } + E_{\bf k} ) t / \hbar } \\
&- \sum_{m} \sum_{ {\bf k} \neq 0, -{\bf q} } 
\sum_{ {\bf l} \neq 0, -{\bf q} } \frac{ v_{\bf k} v_{\bf l} 
u_{ {\bf k} + {\bf q} } u_{ {\bf l} + {\bf q} } V_{\bf q} }
{ E_{ {\bf k} + {\bf q} } + E_{\bf k} + \hbar \omega + i \hbar \eta } 
e^{ -i ( E_{ {\bf k} + {\bf q} } + E_{\bf k} + \hbar \omega + i \hbar \eta ) 
t / \hbar } e^{ i ( E_{ {\bf k} + {\bf q} } + E_{\bf k} ) t / \hbar }
\nonumber
\end{align}
\end{widetext}

\noindent The higher momenta terms (${\bf k} \neq 0, -{\bf q}$) in the
density response function represent processes that are forbidden by
kinematics constraints, and hence they are dropped.

In particular, notice that for {\bf k} = 0 in (\ref{ampl})

\begin{align} \label{m0k0}
\nonumber \langle \Psi_m | a_{\bf q}^\dagger | \Psi_0 \rangle &= 
\langle \Psi_m | ( u_{\bf q} b_{\bf q}^\dagger - v_{\bf q} 
b_{ - {\bf q} } ) | \Psi_0 \rangle \\
&= u_{\bf q} \; \langle \Psi_m | b_{\bf q}^\dagger | 
\Psi_0 \rangle 
\end{align}

\noindent with final state

\beq \label{psim0'}
| \Psi_{m_0'} \rangle = b_{\bf q}^\dagger | \Psi_0 \rangle
\eneq

\noindent and matrix element

\beq \label{matrizm01}
\langle \Psi_m | a_{\bf q}^\dagger | \Psi_0 \rangle = u_{\bf q}
\eneq

\noindent and for {\bf k} = -{\bf q} 

\begin{align} \label{m0k-q}
\nonumber \langle \Psi_m | a_{ - {\bf q} } | \Psi_0 \rangle &= \langle 
\Psi_m | ( u_{ - {\bf q} } b_{ - {\bf q} } - v_{ - {\bf q} } 
b_{\bf q}^\dagger ) | \Psi_0 \rangle \\
&= - v_{\bf q} \; \langle \Psi_m | b_{\bf q}^\dagger | 
\Psi_0 \rangle 
\end{align}

\noindent whose final state is the same state (\ref{psim0'})

The expressions (\ref{m0k0}) and (\ref{m0k-q}) make evident the
meaning of the multiplicative factor $( u_{\bf q} - v_{\bf q} )^2 =
u_{\bf q}^2 + v_{\bf q}^2 - 2 u_{\bf q} v_{\bf q}$ in (\ref{dens}).
$v_{\bf q}^2$ represents the probability of knocking a boson with
momentum $-{\bf q}$ out of the correlated ground state, $u_{\bf q}^2$
represents the probability of knocking into the ground state one of
the boson that are correlated with momentum ${\bf q}$, and $u_{\bf q}
v_{\bf q}$ is the interference between the two processes. It is this
interference factor which provides the suppression of light
scattering\cite{kett2} as a consequence of the rigidity intrinsic of a
superfluid ground state ($v_{\bf q} \neq 0$). Using equations
(\ref{etilde}), (\ref{uv}), (\ref{energy}), and (\ref{sound}) we find,
as ${\bf q} \rightarrow 0$

\beq
\tilde {\epsilon}_{\bf q} = N_0 U_0 ( 1 
+ \frac{ \hbar^2 q^2 }{ 2 m^2 c_s^2 } ) \; ,
\quad 
E_{\bf q} \simeq \hbar q c_s ( 1 + \frac{ \hbar^2 q^2 }
{ 8 m^2 c_s^2 } )
\eneq

\noindent thus 

\beq
\frac{ \tilde {\epsilon}_{\bf q} }{ E_{\bf q} } 
\simeq \frac{ m c_s }{ \hbar q } (1 + 
\frac{ 3 \hbar^2 q^2 }{ 8 m^2 c_s^2 } )
\eneq

\noindent Hence

\beq
u_{\bf q}^2 + v_{\bf q}^2 = \frac{ \tilde {\epsilon}_{\bf q} }
{ E_{\bf q} } \simeq \frac{ m c_s }{ \hbar q } ( 1 + 
\frac{ 3 \hbar^2 q^2 }{ 8 m^2 c_s^2 } )
\eneq

\beq
u_{\bf q} v_{\bf q} = \frac{ 1 }{ 2 } \sqrt{ \frac{ \tilde 
{\epsilon}_{\bf q}^2 }{ E_{\bf q}^2 } - 1 } 
\simeq \frac{ m c_s }{ 2 \hbar q } 
( 1 - \frac{ \hbar^2 q^2 }{ 8 m^2 c_s^2 } )
\eneq

\noindent Putting everything together

\beq
u_{\bf q}^2 + v_{\bf q}^2 - 2 u_{\bf q} v_{\bf q} 
\simeq \frac{ \hbar q }{ 2 m c_s }
\eneq

\noindent This goes to zero linearly as ${\bf q} \rightarrow 0$. The
processes of knocking a particle with momentum ${\bf q}$ out of the
correlated ground state and knocking a particle with momentum $-{\bf q}$
into the ground state both lead to the same final state. That is, the
two processes cannot be differentiated by the scattering process and
thus will interfere quantum mechanically. This is so because neither
of them is an elementary excitation of the system, but a coherent
superposition of them makes a Bogolyubov quasiparticle, which is an
eigenstate of the system. The interference is destructive at small
wavevectors and scattering is suppressed.

We conclude by emphasizing that if there is no repulsion between the
atoms, low energy excitations will not have a sound spectrum, $v_{\bf
k} = 0, u_{\bf k} = 1$ leading to no superfluidity and no suppressed
light scattering. We thus see that suppressed light scattering follows
from the rigidity concomitant to the superfluid state, characterized
by $v_{\bf k} \neq 0$, which is equivalent to a nonzero sound
speed. This rigidity is ultimately due to repulsive interactions
between the bosons. There is no scattering because the ``Principle of
Superfluidity'' forbids it. These are universal properties valid for
any superfluid state which will hold exactly at {\it low enough energy
scales}.

\end{document}